\documentclass{article}
\usepackage{arxiv}
\usepackage[utf8]{inputenc} % allow utf-8 input
\usepackage[T1]{fontenc}    % use 8-bit T1 fonts
\usepackage{hyperref}       % hyperlinks
\usepackage{url}            % simple URL typesetting
\usepackage{booktabs}       % professional-quality tables
\usepackage{amsfonts}       % blackboard math symbols
\usepackage{nicefrac}       % compact symbols for 1/2, etc.
\usepackage{microtype}      % microtypography
\usepackage{cleveref}       % smart cross-referencing
\usepackage{lipsum}         % Can be removed after putting your text content
\usepackage{graphicx}
\usepackage{doi}
\usepackage{listings}
\usepackage{caption}
\usepackage{subcaption}
\usepackage{xcolor}

\colorlet{punct}{red!60!black}
\definecolor{background}{HTML}{EEEEEE}
\definecolor{delim}{RGB}{20,105,176}
\colorlet{numb}{magenta!60!black}

\usepackage{bera}% optional: just to have a nice mono-spaced font

\lstdefinelanguage{json}{
    basicstyle=\normalfont\ttfamily,
    numbers=left,
    numberstyle=\scriptsize,
    stepnumber=1,
    numbersep=8pt,
    showstringspaces=false,
    breaklines=true,
    frame=lines,
    backgroundcolor=\color{background},
    literate=
      {:}{{{\color{punct}{:}}}}{1}
      {,}{{{\color{punct}{,}}}}{1}
      {\{}{{{\color{delim}{\{}}}}{1}
      {\}}{{{\color{delim}{\}}}}}{1}
      {[}{{{\color{delim}{[}}}}{1}
      {]}{{{\color{delim}{]}}}}{1},
}

\title{ColonyOS - A Meta-Operating System for Distributed Computing Across Heterogeneous Platforms}

\author{{\hspace{1mm}Johan Kristiansson} \\
	Department of Computer Science \\
	RISE Research Institutes of Sweden \\
	Luleå, Sweden \\
	\texttt{johan.kristiansson@ri.se} \\
}

\renewcommand{\headeright}{}
\renewcommand{\undertitle}{}

\begin{document}
\maketitle

\begin{abstract}
This paper presents ColonyOS\footnote{https://colonyos.io}, an open-source meta-operating system designed to improve integration and utilization of diverse computing platforms, including IoT, edge, cloud, and HPC. Operating as an overlay, ColonyOS can interface with a wide range of computing environments, fostering creation of so-called \emph{compute continuums}. This makes it possible to develop AI workflows and applications that can operate across platforms. 

At its core, ColonyOS consists of distributed executors that integrate with various underlying platforms based on a distributed microservice architecture. These executors collectively form a \emph{colony}, serving as a unified computing unit. To enable secure integration of various platforms, each colony is provisioned with precisely the resources needed, and all communication is confined within the colony governed by a strict zero-trust security protocol.

Interaction with ColonyOS is done by submitting functional meta-descriptions of computational tasks, called function specifications. These are sent to a Colonies server, which acts as intermediary between applications and the executors. Upon assignment, an executor interprets the meta-description and translates it into an executable format, e.g. a Kubernetes deployment description, a Slurm script, or a direct function call within the executor. Furthermore, a built-in meta-file system enables data synchronization directives to be included in meta-descriptions, enabling seamless data management across platforms.

Ultimately, ColonyOS paves the way for development of hyper-distributed AI applications and workflows, which can seamlessly operate in a computing continuum. The paper describes design principles and implementation details of ColonyOS.
\end{abstract}

\section{Introduction}
Artificial intelligence and machine learning has gained significant traction in recent years. At the same time, development and operation of AI workloads has become increasingly challenging. One difficulty is lack of portability, making it cumbersome to move from one platform to another. Creating and operating fully automated end-to-end workflows across devices, edge, and cloud platforms is even more challenging. 
To address these challenges, the paper presents an open-source meta operating system called ColonyOS, which facilitates execution of computational workloads across a diverse range of platforms. ColonyOS provides a \emph{distributed runtime environment}, called a \emph{colony}, which can effortlessly integrate with any third-party application or other workflow systems.

Developing robust and scalable AI systems is a challenging task that requires deep understanding in several fields. To begin with, an AI model must be trained which requires knowledge in advanced statistics or machine learning. Typically, training and validation data must be pre-processed through various stages before it can be utilized. Although it may be practical for small-scale projects to run the entire training processes on local development computers, larger AI models such as Large Language Models (LLM) typically require access to powerful compute clusters or even High-Performance Computing (HPC) systems. Manual use of such infrastructure can be laborious and time-consuming. Automating the training process enables faster iterations and quicker discovery of useful models.

Taking an AI model into production requires substantial software engineering expertise and collaboration across teams. In contrast to traditional IT workloads, both the data and the model must be managed in addition to the software itself. As most models require regular retraining or re-calibration, it must be possible to update deployed models and software seamlessly without losing information or breaking the system. In many cases, there is a constant flow of data ingested into the system which must be managed even in case of failures. This becomes even more challenging when nodes or parts of the underlying infrastructure become unavailable due to maintenance such as software updates, hardware replacements or sometimes misconfiguration problems.

In some cases, it may be necessary to scale the system to increase or reduce the capacity. This is especially critical when using expensive cloud resources. Scaling the system means that the underlying infrastructure may change at any time, causing instability issues for running services or workflows. Therefore, it must be possible to detect failed computations and re-process failed tasks part of a workflow. Workflows must hence be designed to handle an ever-changing infrastructure, and if a failed computation cannot be restored gracefully, engineers must be able to quickly perform root cause analysis to manually recover the system.

In reality, developing AI applications require integration of multiple systems. For instance, data needs to be captured from an IoT system or pulled from a third-party database running on different domains than the computing cluster itself. With the emergence of edge computing, parts of a data pipeline may also run on edge servers to bring computations closer to data sources. Configuring and setting up such pipelines add even more complexity. 

Additionally, many computing clusters are deployed on-premises. Sometimes it is necessary to temporarily increase the capacity of on-prem clusters by combining resources from multiple providers, for example, adding cloud compute resources to handle peak loads or utilize HPC resources to quickly re-process historical data. Developing hyper distributed workflows where some tasks run in the cloud and others run on HPC systems requires even more software development efforts \cite{wf_challenges} and is beyond the scope of many users, preventing them from utilizing powerful hardware. Clearly, there is a need for a framework that can consolidate various workflow management platforms to simplify development and enable seamless execution across platforms.

The long-term vision of the ColonyOS is to enable uninterrupted access to computing resources, allowing executions and data to flow from one platform to another, creating a compute continuum. The paper suggests that a meta operating system can achieve this vision and support development of hyper-distributed AI applications. The remainder of the paper describes ColonyOS and how it can be used to create robust and scalable AI workflows.

\section{Related work}
To promote innovation of European cloud technologies, the European Commission is currently actively funding several meta-operating system Horizon projects, including aeROS \cite{aeros}, FluiDOS \cite{fluidos}, ICOS \cite{icos}, NebulOuS \cite{nebulous}, NEMO \cite{nemo}, NEPHELE \cite{nephele}. 
These efforts highlight the critical role that meta-operating systems could play in facilitating hyper-distributed applications across computing landscapes.

The concept of a meta-operating system is not new with the Robot Operating System (ROS) \cite{ros} being an early example. Although ROS focuses on robotics, its design and features align with meta operating system principles, e.g. abstracting hardware and software complexities to streamline application development and deployment across devices. More recently, the RAMOS \cite{34962672} meta operating system was developed to transform cloud-edge-IoT systems into a dynamic, peer-to-peer network focused on context-aware machine learning and energy efficiency. Debab et al. \cite{debab2018boosting} also proposed a Meta-OS that uses heterogeneous monolithic kernels to improve reliability and performance of web services.

While many of the meta-operating system initiatives build on Kubernetes or strive to provide holistic full-stack solutions that cover everything from resource optimization to sophisticated network virtualization, ColonyOS takes a different approach. Instead of offering a full-stack solution, ColonyOS primarily aims to provide a \emph{simple architecture to facilitate integration with other platforms}. From this perspective, ColonyOS could even be combined with other meta operating system projects, e.g. be integrated with FluiDOS to more efficiently handle edge resources.

ColonyOS is a unique combination of several technologies, including distributed runtimes, orchestration, workflow management, grid computing, zero-trust security, and data synchronization. The subsequent text examines ColonyOS in relation to some of these core technologies.

\subsection{Workflow management}
Distributed runtimes enable execution of software applications on many interconnected computing resources. It provides mechanisms and services to distribute, manage, and coordinate the execution of tasks. Apache Spark \cite{spark} and Apache Flink \cite{flink} are well known examples of distributed runtimes. Apache Spark is specifically designed for distributed data processing and analytics. It operates on distributed data sets, performing transformations and computations on those datasets. It offers a rich set of APIs and libraries for data processing. Apache Flink, on the other hand, is a distributed stream processing framework to process continuous streams of data in real-time. Flink is based on a dataflow programming model and is commonly used in real-time data analytics applications. Unlike Spark and Flink, which primarily operate on distributed data sets or data streams, ColonyOS emphasizes coordination and integration across platforms. ColonyOS aims to tackle challenges related to portability, automation, scalability, and security in heterogeneous computing environments.

Workflow management has been extensively studied in both academic and industrial settings with numerous approaches \cite{scafe, synapse, service_wfs, schmitt2022workflow, GarciaRepresa1740746, Ouyang2010, NIKOLOV2021100440, workflow_in_bigdata, argowf} proposed to address the challenges in this field. Recently, Apache Airflow \cite{apache_airflow} has become a popular open-source workflow management system for handling data engineering pipelines. 

Similar to ColonyOS, Apache Airflow enables developers to create custom operators and executors that can be integrated with various systems. Additionally, Apache Airflow offers an HTTP API that makes it possible to develop Software Development Kits (SDKs) in various programming languages. Apache Airflow could be used internally in ColonyOS. However, Apache Airflow must typically be integrated with a message broker, such as RabbitMQ \cite{rabbitmq} or Kafka \cite{apache_kafka} to implement task queues, resulting in a more complex architecture. 

Today, utilization of serverless computing is experiencing a significant growth \cite{cognit}, primarily due to its potential to liberate developers from the burden of managing underlying cloud infrastructures. Attempts are currently being made to implement serverless workflow management systems. For example, OpenWolf \cite{openwolf} is a serverless workflow engine designed to utilize the Function-as-a-Service (FaaS) paradigm to compose complex scientific workflows. It is based on OpenFaaS \cite{openfaas}, which allows functions to run on Kubernetes clusters. ColonyOS can also be used to implement serverless workflow management systems. However, in contrast to previous work, ColonyOS is designed to be platform independent and does not require a container orchestration platform like Kubernetes. By using a zero-trust security protocol, functions can be securely executed by distributed executors deployed anywhere on the Internet. 

\subsection{Microservices}
ColonyOS is based on a distributed microservice architecture, which makes it suitable for DevOps software development. Currently, microservices are primarily used to implement large-scale and high-availability web applications. It has not yet become a prevalent design principle for workload management or implementation of HPC applications. Instead, simple job scripts are commonly used. J. Represa et al. \cite{GarciaRepresa1740746, GarciaRepresa1640771} explore various challenges associated with developing microservice-based workflow management for industrial automation within the context of the Arrowhead project \cite{delsing2017iot}. The authors conclude that microservice-based workflow technologies are viable for industrial applications, particularly due to their inherent flexibility. 

\subsection{Grid computing}
Grid computing \cite{grid_computing} is a distributed computing model that allows multiple computers, which may be geographically dispersed, to collaborate in addressing large-scale computational challenges. In grid computing, the role of brokers is crucial for coordinating tasks and managing resources efficiently across the network. ColonyOS can be regarded as a modern grid computing platform. However, unlike traditional grid computing platforms \cite{globus}, which typically seeks to establish a global network of computational resources, ColonyOS is more application-specific, where each application operates within its own dedicated grid, tailoring the distributed computing environment to its unique requirements and workflows.

The primary contribution of this paper is a technical description of a meta-operating system designed to support development of hyper-distributed applications and workflows running across diverse platforms. ColonyOS is intentionally designed to be simple and platform-independent, ensuring broad applicability without requiring extensive integration with specific platforms like Kubernetes. Instead, it maintains the flexibility to interface with a variety of platforms while simultaneously enhancing the resilience of applications in the event that one platform becomes temporarily unavailable. The following sections describe the design principles and implementation details of ColonyOS.

\begin{figure}[t]
	\centering
    \includegraphics[scale=0.50]{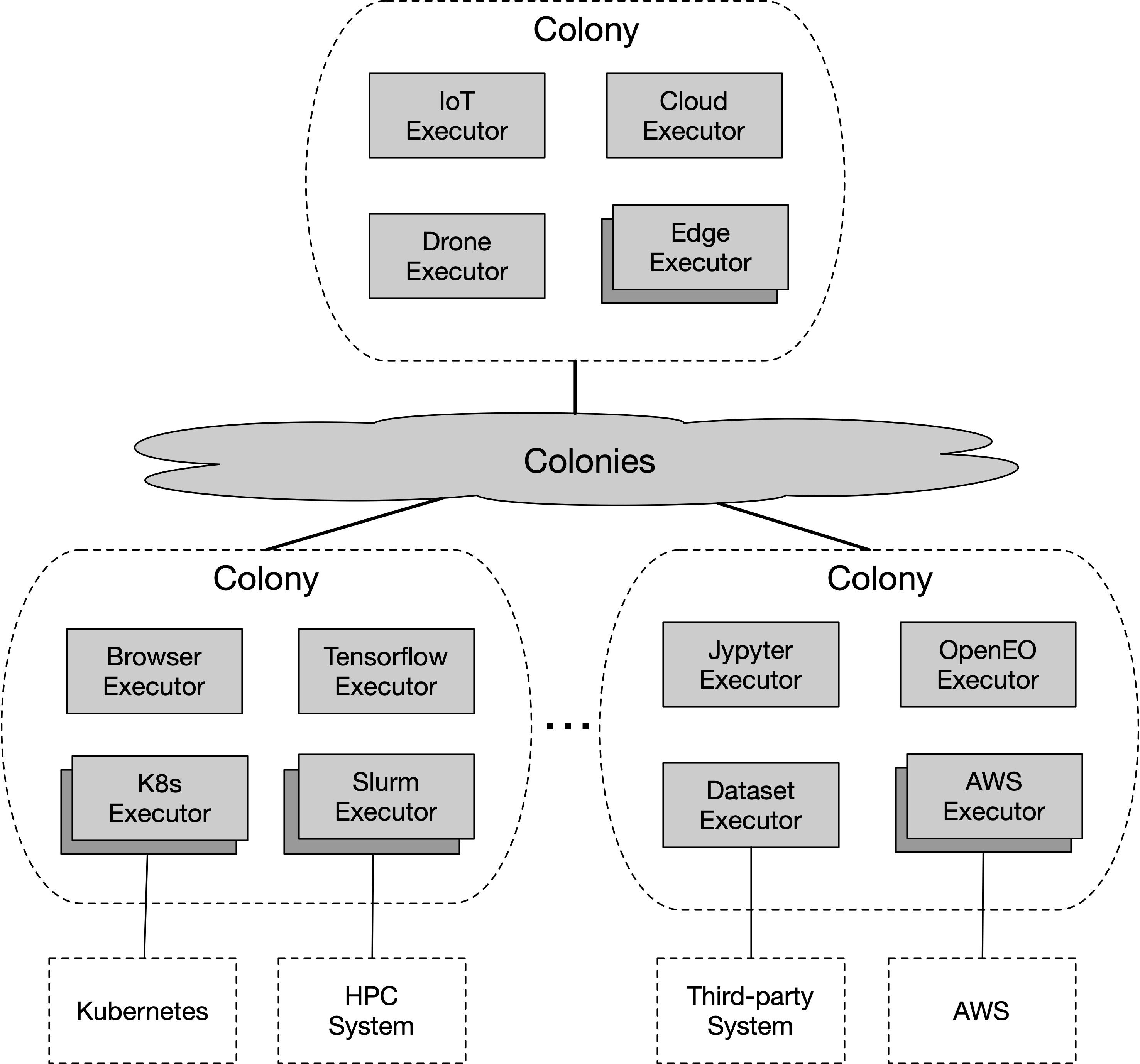}
	\caption{Overview of ColonyOS. Executors may be deployed anywhere on the Internet.}
	\label{fig:overview}
\end{figure}

\section{ColonyOS}
\label{sec:headings}
A core concept of ColonyOS, and many other operating systems, is the notion of \emph{processes}. A process in ColonyOS contains meta-information about computations executed by remote computer programs called \emph{executors}. Specifically, a process contains a definition of a function to be executed as well as contextual information such as execution status and the result of the computation. It is essential to point out that a process does not necessarily have to be an operating system process, but can represent any type of computation, e.g. a remote procedure call executed by any kind of software. 

Figure 1 depicts an overview of the components part of ColonyOS. The \emph{Colonies servers} form the backbone of the framework, functioning as a centralized control system for task submission and assignment. The Colonies server serves as a job broker for executors, almost like an employment agency for people. All execution information and history are stored in a database maintained by the Colonies servers. Upon submission, a process is stored in the database, serving as a queueing system. When a process is assigned to an executor its execution status is changed from \emph{waiting} to \emph{running}. Consequently, it is possible to submit a process that should run in the future even if no executors can run the process at the moment. In this way, ColonyOS supports both batch and real-time processing. 

\emph{Executors} are responsible for executing processes assigned by the Colonies servers. Several executors form a \emph{colony}, which is a collection of executors operating as a cohesive unit where each executor is responsible for executing specific types of processes. One could view a Colony as an organization (like a company) of distributed computer softwares acting as a single virtual computer system. To interact with other executors, executors must prove their colony membership using a cryptographic protocol that follows the zero-trust security principle: \emph{never trust, always verify}. This security protocol ensures that users can keep control even when executors are distributed across platforms. Zero-trust security is fundamental to ColonyOS and will be further discussed in Section \ref{zerotrustsecurity}. The following section outlines the underlying design principles and describes ColonyOS in more detail. 

\subsection{Microservices}
Microservices \cite{microservices} is an architectural design pattern in which an application is structured as a collection of small, independently deployable, and loosely coupled services that communicate with other microservices through a well-defined API. By dividing the application into smaller, focused microservices, applications become easier to understand, maintain, and develop. In ColonyOS, executors are microservices having the following characteristics:

\begin{itemize}
\item \textbf{Single responsibility:} Each executor is only accountable for executing specific functions. This makes the system easier to understand, develop, test and maintain.  
\item \textbf{Loosely coupled:} Executors are designed with minimal dependencies on other executors, enabling different software development teams to work independently. For instance, a data engineering team may handle the implementation of pre-processing functions, while a data science team oversees machine learning functions, and another team manages visualization or customer integration etc.

\item \textbf{Scalability:} Executors can be deployed independently and be scaled horizontally. This enables efficient resource utilization, parallelism, and improved performance. 

\item \textbf{Resilience:} The failure of a single executor does not compromise the entire application or workflow. Executors' isolation from one another contributes to a more resilient and fault-tolerant system. If an executor crashes during execution, the process is automatically re-assigned to another executor.

\item \textbf{DevOps and Continuous Integration:} Executors' inherent fault-tolerant design permits changes to individual executors without impacting the entire system. This makes it possible to seamlessly update the system. 
\item \textbf{Decentralized governance:} As executors are technology-agnostic, different software development teams can make independent technology and design decisions when developing new executors, promoting greater flexibility and adaptability. For example, some executors may be implemented in Rust, while others may use Python to leverage state-of-the-art machine learning frameworks.
\end{itemize}

Although microservices have attractive software engineering properties, they also introduce additional complexity that must be managed by ColonyOS. To implement a workflow management framework supporting distributed microservices, the following challenges must be addressed:

\begin{itemize}
\item \textbf{Process management:} ColonyOS must be able to distribute processes among available executors based on their capabilities and current workload. This involves assigning processes to the most appropriate executor and then monitoring execution progress.

\item \textbf{Load balancing:} ColonyOS must manage load balancing among executors to optimize the overall performance.

\item \textbf{Fault tolerance and recovery:} In the event of executor failures, ColonyOS must be able to re-assign processes to other executors, manage executor restarts, or trigger recovery mechanisms to maintain system reliability and resilience. In addition, ColonyOS must be able to handle re-starts or crashes of Colonies servers and prevent any internal states from becoming corrupted. 

\item \textbf{Service discovery:} ColonyOS must enable executors to dynamically register and deregister from Colonies servers. It must be possible to deploy executors anywhere on the Internet, even behind firewalls.

\item \textbf{Workflow orchestration:} ColonyOS must coordinate and orchestrate complex multi-step workflows executed by several executors, sometimes in parallel where executors run on different platforms. This includes defining the sequence in which processes should be executed, manage dependencies among processes, and ensure that workflows execute successfully to completion.

\item \textbf{Monitoring and debugging:} ColonyOS must monitor the overall system, allowing administrators to track the health, performance, and resource utilization of the executors and the system as a whole.
\end{itemize}

As will be discussed, ColonyOS uses a combination of technologies to address the aforementioned challenges. A fundamental design principle is statelessness, which makes ColonyOS simpler, more reliable and scalable, and easier to implement. To ensure reliability and data consistency, a distributed consensus algorithm is needed to handle high-availability and server crashes. The remainder of this section describes how ColonyOS is implemented. The next discusses the role of queues to support batch jobs and realtime processing, but also discusses queues roles in implementing loosely coupled systems.

\begin{figure}
     \centering
     \begin{subfigure}[b]{0.4\textwidth}
         \centering
         \includegraphics[scale=0.45]{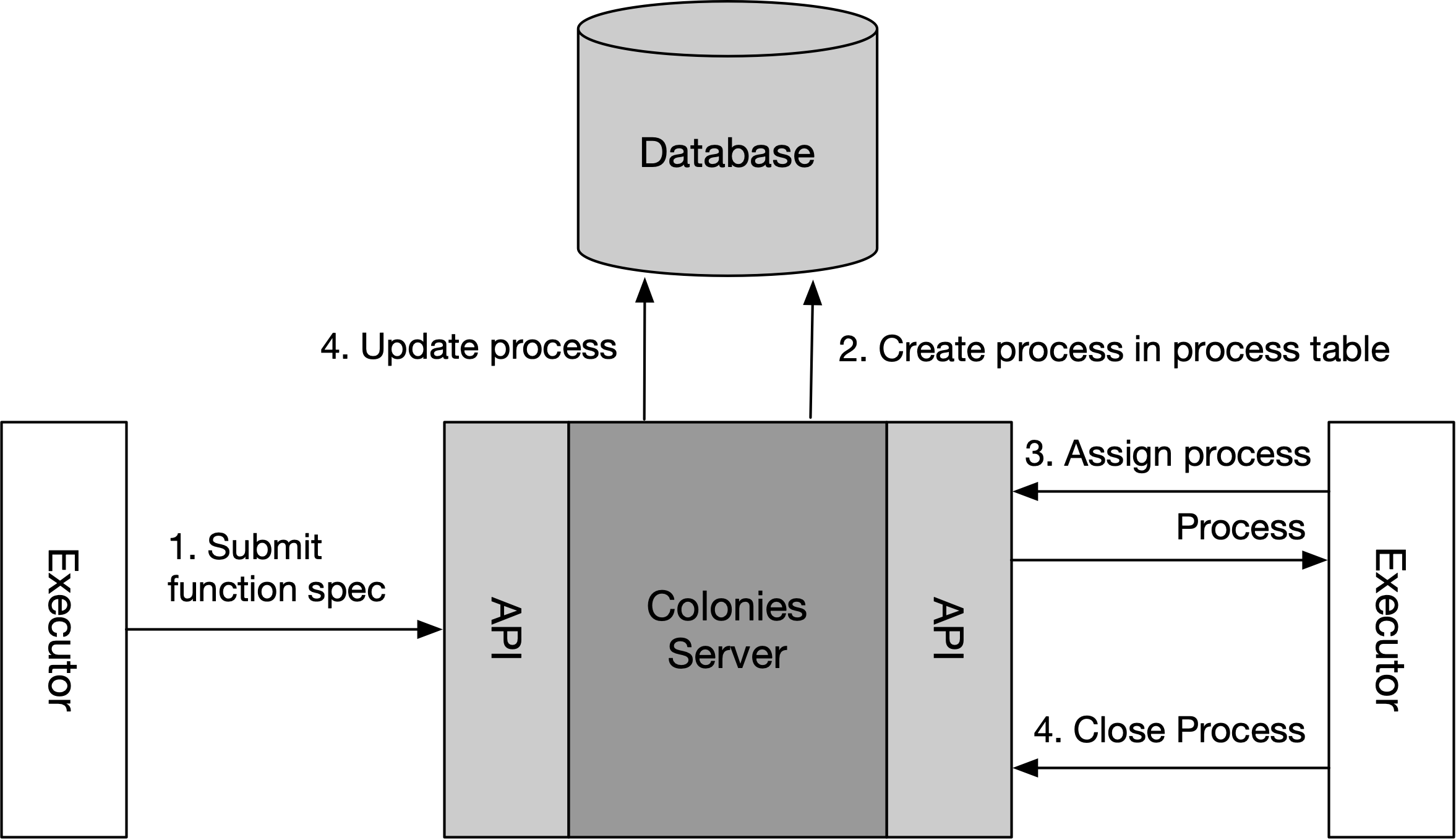}
         \caption{Process assignment steps.}
     \end{subfigure}
     \hfill
     \begin{subfigure}[b]{0.4\textwidth}
         \centering
         \includegraphics[scale=0.45]{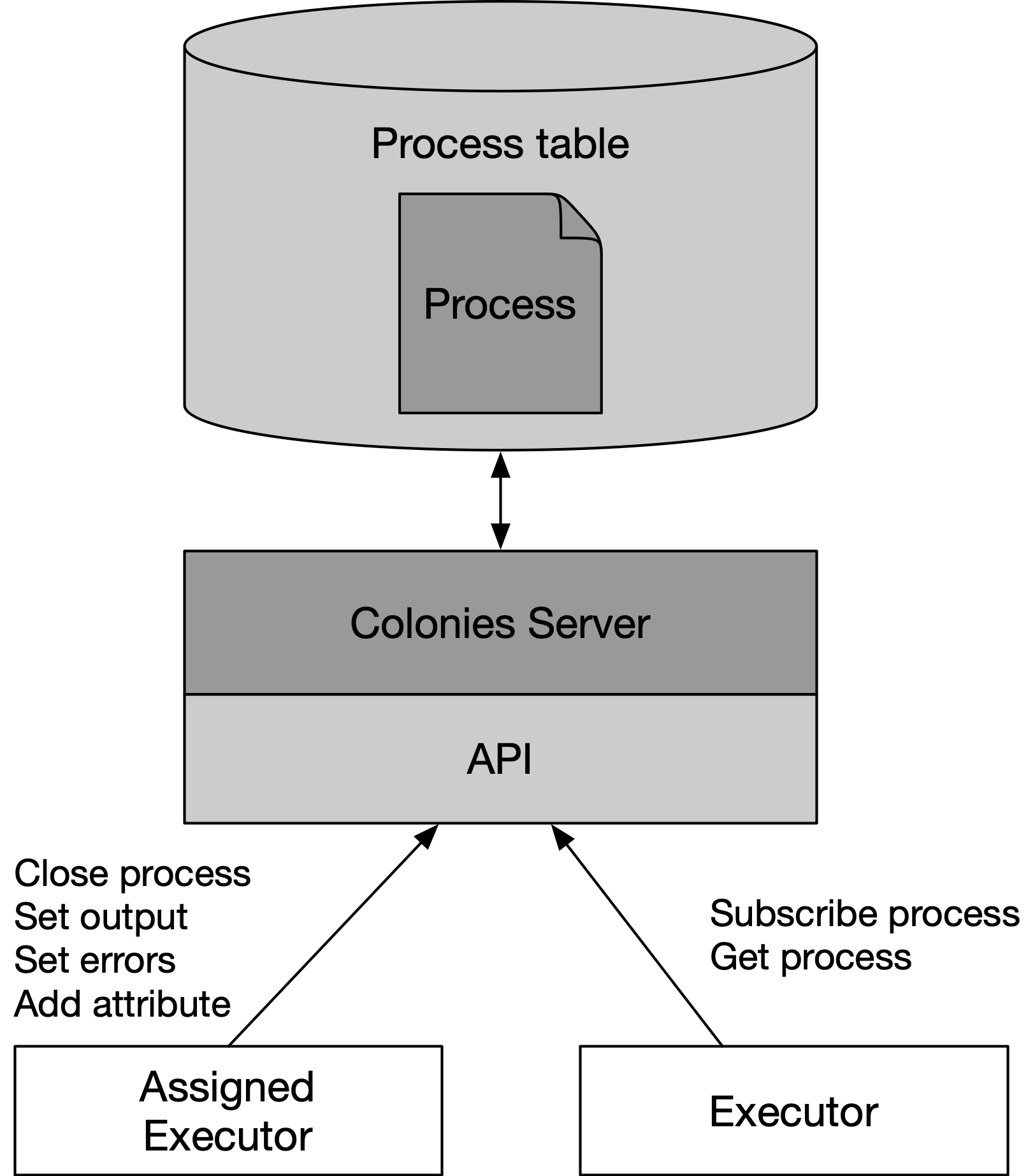}
         \caption{Manipulation of a process.}
     \end{subfigure}
     \caption{Process management via ColonyOS HTTP API. Note that only the assigned executor has write access to the process database entry.}
     \label{fig:process_assignment}
\end{figure}

\subsection{The role of queues as seperation of concerns}
Separation of concerns (SoC) is a design principle to break down a complex software system into smaller, more manageable parts. For example, HTTP APIs can be used to abstract away implementation detail and provide a clear and simple interface for interacting with a particular service. However, APIs alone are insufficient to handle dynamic environments where components frequently fail or the underlying infrastructure is constantly changing. To address such environments, an alternative mechanism is needed. 

Queues enable software services to communicate indirectly by acting as a buffer between them. Queues make it possible to decouple executors and make them operate independently, for example an executor can be updated or replaced without affecting other executors. Queues also enable asynchronous communication between executors, enabling them to execute tasks at their own pace. This ensures that slower executors do not bottleneck faster ones, leading to a more efficient and scalable system. Most importantly, queues enable load balancing by distributing tasks among executors, thus making it possible to parallelize workflow execution. 

Queues can be implemented in different ways. While message brokers are a common solution, ColonyOS adopts an alternative strategy and is based on a standard database. One key advantage of this approach is that it enables fine-grained process assignments, making it possible to assign specific processes to particular executors. For instance, an executor can be limited to only execute processes capable of running in web browsers. This level of granularity cannot easily be implemented using message brokers which generally do not allow introspection of queues, or provide the ability to pull specific messages out of the queue. Generally, the only way to retrieve a specific message is to pull all messages from the queue, obtain the message, and then place all remaining messages back into the queue in the same order. In contrast, a database can function as a queue and a query can match any columns making it possible to assign an executors to specific processes.

\begin{table}[t]
	\caption{Process Table}
	\centering
	\begin{tabular}{llllll}
		\toprule
		\cmidrule(r){1-2}
        Process Id & Function Spec & Wait for Parents & Executor  & State      & Priority Time \\
		\midrule
        $P_{1}$    & $F_{1}$       & $False$          & $E_{1}$   & Successful & 1679906715352024000 \\
        $P_{2}$    & $F_{2}$       & $False$          & $E_{1}$   & Running    & 1679906715353453000 \\
        $P_{3}$    & $F_{3}$       & $False$          & $E_{2}$   & Running    & 1679906715354286000 \\
        $P_{4}$    & $F_{4}$       & $True$           & -         & Waiting    & 1679906715355188000 \\
		\bottomrule
	\end{tabular}
	\label{proctable}
\end{table}

\begin{table}[t]
	\caption{Function Specifications}
	\centering
	\begin{tabular}{llllll}
		\toprule
		\cmidrule(r){1-2}
        Function Spec & Function        & Executor Type & Priority & Max Exec Time & Max Retries \\
		\midrule
        $F_{1}$       & gen\_nums()     & Edge          & 1        & 200 s         & 5 \\
        $F_{2}$       & square()        & Cloud         & 1        & 200 s         & 5 \\
        $F_{3}$       & square()        & Cloud         & 1        & 200 s         & 5 \\
        $F_{4}$       & sum()           & Browser       & 1        & 200 s         & 5 \\
		\bottomrule
	\end{tabular}
	\label{functable}
\end{table}

\subsection{Process tables}
ColonyOS enables executors to interact with each other by submitting function specifications to the Colonies servers. Once submitted, other executors can connect to the servers to receive process execution assignments. A new process entry is automatically added to the process table database when a function specification is submitted to the Colonies server.

When an executor connects to the Colonies server, the server hangs the incoming HTTP connection\footnote{An alternative protocol is to use WebSockets or gPRC to communicate with the Colonies server.} until the executor is assigned a process, or until a connection timer expires. The Colonies server never connects to the executors. Rather, it is the responsibility of the executors to connect to the Colonies server. This enables executors to be deployed anywhere on the Internet, behind firewalls, commercial telco networks, or even in web browsers.

Figure \ref{fig:process_assignment} depicts an executor submitting a function specification that is later assigned to another executor. When registering, executors have to specify to the Colonies server which functions they are capable of executing. The Colonies server then makes sure that function specifications match the capability of executors. 

\begin{equation}
    \label{eq:pt}
    priority_{time}=submission_{time} - priority \cdot 10^9 \cdot 60 \cdot 60 \cdot 24
\end{equation}

Table \ref{proctable} shows an example of a process table. The process table also contains a reference to a function specification depicted in Table \ref{functable}. To assign a process to an executor, the Colonies server searches in the process table to find a process matching a waiting executor. By using the \emph{priority time} column to sort the processes, the process table can function as a queue. This can be done in SQL by utilizing the \emph{order by} clause to sort processes according to their submission time. 

\begin{lstlisting}[basicstyle=\small, label=fig:function_spec, language=json, basicstyle=\small, caption=Example of a function specification.]
{
   "conditions": {
       "colonyid": "0c1168fe986ffe39fad14f17e0bd9e5896f6d968405ac0fb3380154109ee4022",
       "executortype": "helloworld_executor"
   },
   "funcname": "helloworld",
   "args": ["hello world"],
   "maxwaittime": 10,
   "maxexectime": 100,
   "maxretries": 3,
   "priority": 1
}
\end{lstlisting}

To make it possible to handle priorities, the submission time is adjusted so that higher priority processes are processed before lower priority processes. When a process is submitted, a \emph{priority time} value is calculated and stored in the process table. Equation \ref{eq:pt} shows how the priority time is calculated for a nanosecond timestamp. 

\begin{figure}[h]
	\centering
    \includegraphics[scale=0.5]{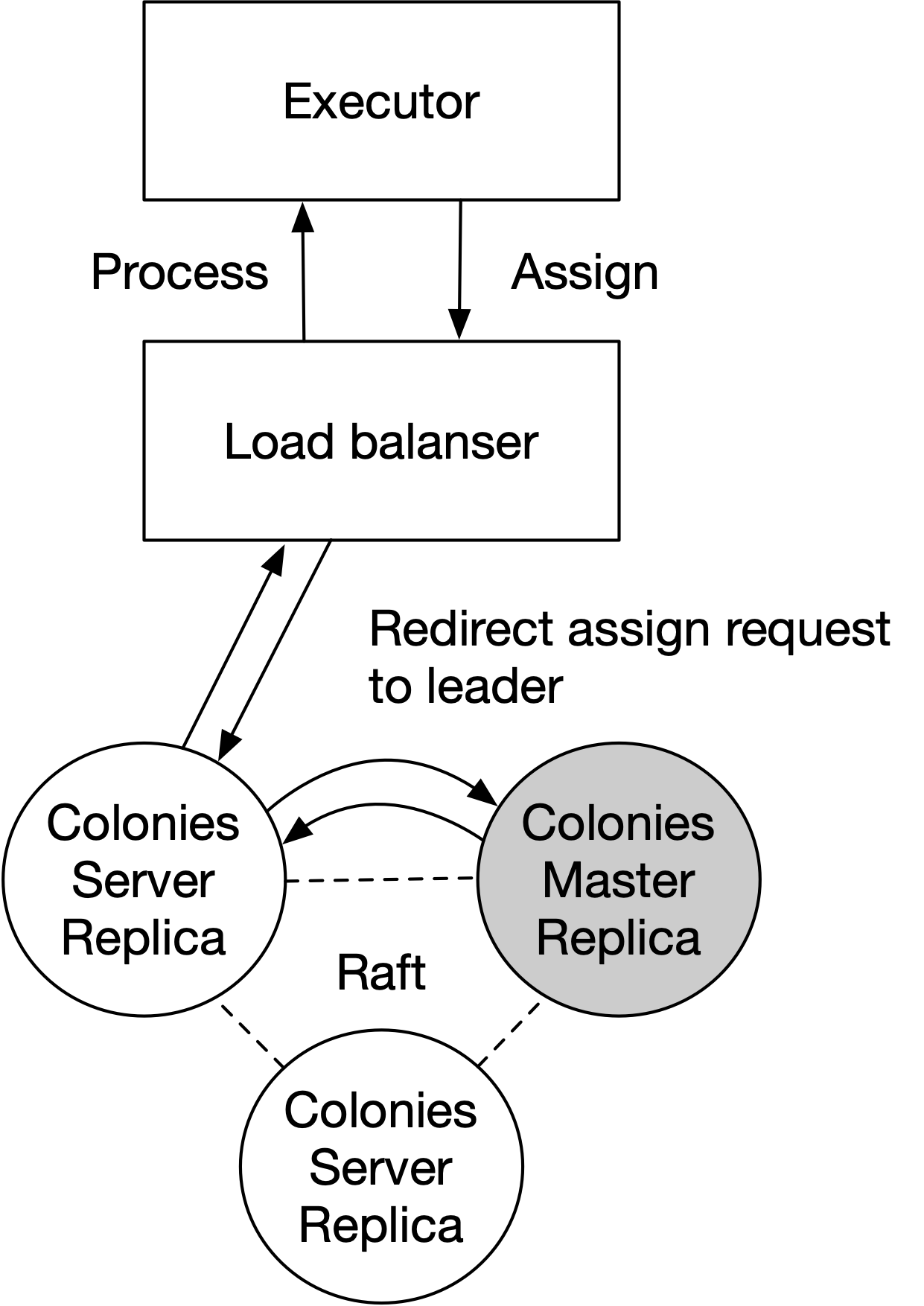}
	\caption{High-availability deployment.}
	\label{fig:ha_deployment}
\end{figure}

\subsection{A stateless failsafe mechanism}
\label{stateless_error_management}
The primary objective of all ColonyOS API requests is to alter some state stored in the database or retrieve information from the database. ColonyOS is designed to be stateless, meaning that Colonies servers do not keep any information in memory between requests. Each request is handled independently without relying on information from previous requests. 

Figure \ref{fig:function_spec} shows an example of a function specification. The \emph{maxexectime} attribute specifies the maximum time an executor may run a process (in this case, 100 seconds). Before a process is assigned to an executor, the Colonies server updates the process entry in the process table database and calculates a deadline when the process must be finished. The server then regularly checks for any running processes that have exceeded their deadlines. If such a process is detected, it is reset, allowing it to be re-assigned to other executors. 

Making it possible to specify maximum execution time is a simple but powerful mechanism. To scale up a system, more executors can simply be deployed. Scaling down, however, can be more challenging. One solution is to select a set of executors to be removed and then starve them out by denying them new process assignments. Another, simpler solution, is to immediately destroy the executors and use the \emph{maxexectime} failsafe mechanism to move back processes from defunct executors to the queue. The \emph{maxexectime} failsafe mechanism ensures that processes will eventually be executed even in the case of failures. This mechanism also relieves the burden of the user to check if a process has been executed or not, as they can simply look up the process in the database to get its current status. 

Utilizing the \emph{maxexectime} failsafe mechanism not only enhances system reliability, but also provides an opportunity to apply Chaos engineering \cite{chaos_engineering}. For example, a Chaos monkey can be used to deliberately terminate executors. If executors are deployed on Kubernetes, Kubernetes will then automatically redeploy terminated executors. The constant flux of executor replacements ensures that the system is capable of gracefully tolerating failures.

\subsubsection{Data consistency and distributed consensus}
Synchronization is essential to prevent data inconsistency and race conditions when accessing shared resources concurrently with multiple threads. However, synchronization can also slow down execution as only one thread can access critical sections at a time. By carefully designing multithreaded applications and using right synchronization techniques, it is possible to minimize the performance impact while still ensuring data consistency and correctness.  

The \emph{assign} API request binds a process from the database to an executor. Given the multi-threaded nature of the Colonies server, it is essential that the \emph{assign} request is synchronized to ensure that only one thread at a time can modify the database and update the process table, thus preventing multiple executors from being assigned to the same process. To ensure that only one executor can be assigned to a process, the \emph{assign} request must be synchronized. It is worth noting that synchronization is not necessary for other requests. For example, as the \emph{submit} request only adds new entries to the process table, hence there are no race conditions. The \emph{close} request sets the output of the function innovation and updates the process state to either successful or failed in the process table. Since there can only be one executor assigned to a process there are no race conditions and consequently no need for synchronization.

To minimize downtime, ColonyOS supports high-availability deployments. If one Colonies server crashes, an executor simply needs to resend the failed request, which will then be served by another Colonies server replica. However, by introducing multiple Colonies servers, there is again a risk of race conditions when assigning processes to executors. This means that ColonyOS must coordinate which replica server should serve incoming assign requests so that precisely one executor is assigned to a particular process.

Raft \cite{raft} is a consensus algorithm specifically designed to manage a replicated log within a distributed system. Raft makes it possible to implement distributed leader elections where a single server takes on the role of leader while the remaining servers act as followers. The leader is responsible for managing the replicated log, processing client requests, and replicating entries to the followers. Followers passively replicate the leader's log and participate in leader elections. 

By using the Raft protocol, ColonyOS can redirect incoming \emph{assign} requests to the Colonies server replica elected as the leader, thereby ensuring that only one Colonies server replica handles assign requests. A new leader is automatically elected in the event that a Colonies server replica fails. Figure \ref{fig:ha_deployment} shows an overview of a high-availability deployment. 

\begin{table}[h]
	\caption{Dependency Table}
	\centering
	\begin{tabular}{lll}
		\toprule
		\cmidrule(r){1-2}
        Process Id & Name       & Dependencies           \\
		\midrule
        $P_{1}$    & $Task_{1}$ & -                      \\
        $P_{2}$    & $Task_{2}$ & $Task_{1}$             \\
        $P_{3}$    & $Task_{3}$ & $Task_{1}$             \\
        $P_{4}$    & $Task_{4}$ & $Task_{2}$, $Task_{3}$ \\
		\bottomrule
	\end{tabular}
	\label{deptable}
\end{table}

\subsubsection{Workflows}
A workflow is a series of processes that need to be completed in a specific order. In ColonyOS, workflows are represented as directed acyclic graphs (DAGs) where nodes represent processes and edges represent dependencies and data flow between processes. Like processes, workflows are managed completely stateless. When a workflow is submitted, all processes part of the workflow are submitted and added to the process table similar to how ordinary processes are handled. To control the order processes are assigned, ColonyOS sets a flag, \emph{wait for parents} to prevent processes to be assigned to an executor before their dependencies have been satisfied. Note that processes may run in parallel if a sufficient number of executors are available, assuming their parent processes have finished.

\begin{figure}[h]
	\centering
    \includegraphics[scale=0.30]{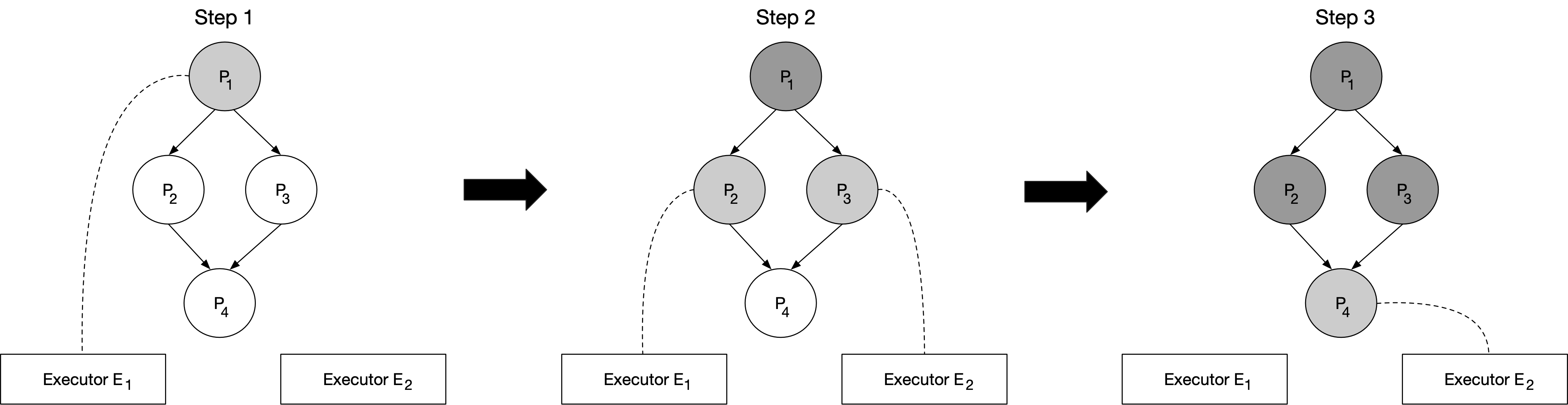}
	\caption{Workflow execution timeline.}
	\label{fig:workflowexec}
\end{figure}

When a process is closed, the Colonies server checks its child processes to see if all their parent processes have also completed. If all parents of a child process have finished, the server updates the process table in the database by setting the flag \emph{wait for parents} to \emph{false}, which allows child processes to be assigned to executors. This operation is also done statelessly when processing the \emph{close} request. 

Figure \ref{fig:workflowexec} shows an execution timeline for a workflow. Upon submission, process \(P_{1}\) is assigned to Executor \(E_{1}\). After being closed, processes \(P_{2}\) and \(P_{3}\) are simultaneously assigned to \(E_{1}\) and \(E_{2}\), allowing for concurrent execution. Lastly, process \(P_{4}\) is assigned to \(E_{2}\).  

To be able to generate a DAG, an additional database table is required. Table \ref{deptable} shows an example of a dependency table storing relationships between processes. The table is updated when a workflow is submitted to a Colonies server. As mentioned before, when an executor is assigned a process, it obtains exclusive rights to that process within the process table. This exclusivity also means that it is possible for an executor to dynamically add new children processes to the workflow DAG on the fly. As a result, it becomes possible to implement data processing patterns like MapReduce \cite{mapreduce} similar to Hadoop \cite{hadoop}.  

\begin{table}[h]
	\caption{Input/Output}
	\centering
	\begin{tabular}{lll}
		\toprule
		\cmidrule(r){1-2}
        Process Id & Input & Output \\
		\midrule
        $P_{1}$    & & [2,3] \\
        $P_{2}$    & 2 & 4 \\
        $P_{3}$    & 3 & 9 \\
        $P_{4}$    & [4,9] & 13 \\
		\bottomrule
	\end{tabular}
	\label{inouttable}
\end{table}

To manage data flow between processes that are part of a workflow, it is necessary to store the input and output values from function executions in the database. An example of such a database is presented in Table \ref{inouttable}. An executor can subsequently query the output from a parent process and use that as arguments while invoking a function.

\subsubsection{Cron}
ColonyOS is based on a stateless architecture where each request is treated as an independent transaction which is processed without any information from prior requests. Since there is no session data stored in memory, it becomes possible to easily switch between servers without disruptions or data losses. In a stateful architecture on the other hand, the server must maintain a record of connected clients' state and session data in memory, which can result in complications if a server fails or crashes. 

Cron is a time-based job scheduler designed to trigger workflows at predefined intervals. It can for example be used to automate tasks such as routinely fetching data from external systems. To make ColonyOS robust and easy to implement, it is essential that Cron workflows also are managed statelessly. This can be achieved by introducing an additional database table to store information about Cron workflow, and let the elected Colonies server replica leader assume the responsibility of managing Cron workflows through a two-step process. 

Firstly, the Colonies server replica leader calculates a future deadline, indicating when a specific Cron workflow is scheduled to be executed. The leader server then conducts periodic scans of the Cron table. When the current time surpasses the calculated deadline for a particular workflow, a new workflow is automatically submitted. Note that this protocol is completely stateless as no session information is stored in memory between cron scans. In the event that the leader Colonies server crashes, a new leader is selected and assumes the responsibility of managing Cron workflows according to this two-step protocol.

\subsubsection{Generators}
Another feature of the ColonyOS is so-called generators. Generators are workflow templates that are automatically triggered when specific conditions are met. Generators can be used to accumulate input data into bulks to improve performance or parallelism. For example, an external service downloads satellite images and can use a generator to automatically triggered a workflow when 10 images are collected. This is achieved by sending a special \emph{pack} request containing input data to a Colonies server, targeting a certain generator. 

Generators facilitate integration with third-party systems as it allows for stateless implementation where interactions with ColonyOS can be done in a fire-and-forget style. ColonyOS guarantees that all pack requests submitted to a generator are processed correctly, even if a Colonies server is restarted or fails. Without this robustness guarantee, third-party integration software would have to use an alternative method such as a message broker or a database to enqueue input data and periodically empty the queues to generate workflows themselves. This adds complexity to the integration process and may require additional infrastructure components. Generators are a powerful feature of ColonyOS that can help simplify integration and implement robust data processing pipelines.

Generators are implemented using a similar method like how Cron workflows are implemented. The elected leader Colonies server takes on the responsibility for scanning a special generator table to decide if a workflow should be triggered. This approach ensures that there is a single point of control for generating workflows and prevents generation of duplicate workflows. If the leader server crashes, a new leader Colonies server is automatically elected, which takes over the responsibility for triggering generators. This ensures that the generation of workflows is not impacted by server failures, and the system can continue to operate smoothly even in case of failures. It can be worth pointing out that since the \emph{pack} request only adds information to the generator table and does not manipulate any states, it can be processed by any Colonies server without synchronization precautious. This means that any Colonies server in the cluster can handle incoming \emph{pack} requests, which increases the scalability and fault-tolerance of the system as a whole.

\subsubsection{Meta-filesystem}
ColonyOS features a meta-filesystem called the Colony Filesystem (CFS). Unlike a traditional filesystem, CFS does not store actual files but rather stores metadata about files. This metadata includes information such as file names, checksums, Internet addresses of servers from which to fetch or upload data, details about the protocols used, and optionally, credentials for accessing the files. CFS is designed to support any type of protocol and storage technology, for example S3 or the InterPlanetary File System (IPFS). 

CFS is integrated into ColonyOS by introducing a new table in the Colonies database for storing file metadata. Executors can use CFS for data synchronization. Within function specifications, it is possible to define files for synchronization, allowing executors to automatically download data needed to execute a process and upload results upon process completion. An example of such a function specification, including CFS metadata, is illustrated in Figure \ref{fig:cfs_func_spec}.

\begin{lstlisting}[basicstyle=\small, label=fig:cfs_func_spec, language=json, basicstyle=\small, caption=Example of a function specification.]
{
    "conditions": {
        "executortype": "ice-kubeexecutor",
        "nodes": 1,
        "processes-per-node": 1,
        "mem": "1Gi",
        "cpu": "500m",
        "walltime": 200,
        "gpu": {
            "count": 0
        }
    },
    "funcname": "execute",
    "kwargs": {
        "cmd": "python3",
        "docker-image": "python:3.12-rc-bookworm",
        "rebuild-image": false,
        "args": [
            "/cfs/src/{processid}/helloworld.py"
        ],
        "keep_snapshots": false
    },
    "fs": {
        "mount": "/cfs",
        "snapshots": [
            {
                "snapshotid": "{snapshotid}",
                "label": "/src",
                "dir": "/src/{processid}",
                "keepfiles": false,
                "keepsnaphot": false
            }
        ]
    },
    "maxwaittime": -1,
    "maxexectime": 100,
    "maxretries": 3
}
\end{lstlisting}

In CFS, all files are immutable, meaning they cannot be altered but only completely replaced. This approach simplifies caching of files, as the immutability guarantees the consistency and reliability of the data being cached. Since files cannot be altered once created, this also reduces the risk of data tampering and unauthorized modifications. Immutability ensures that data remains in its original, unaltered state, and thus provides a more secure environment where data integrity can be maintained over time.

Another reason for immutability is prevention of race conditions. In computing, a race condition occurs when multiple processes access and manipulate the same data concurrently, leading to inconsistent or unexpected results. Since immutable files cannot be modified once they are created, this effectively eliminates the possibility of race conditions related to file modifications. This ensures that when a file is read, it cannot be changed by another process at the same time, leading to more predictable and reliable system behavior. By using immutable files, CFS enables stable and consistent data handling across various executors.

Additionally, CFS offers functionality to create snapshots. This feature is essential for generating immutable copies of whole directory trees, which can include source code or data intended for execution or processing. Since processes can be queued and may have to wait before execution, snapshots ensure that the referenced data remains unaltered and that files cannot be added or removed while the process is in the queue, awaiting execution.

\subsubsection{Zero-trust security}
\label{zerotrustsecurity}
As executors may be deployed across multiple platforms, there is a need for a unified security model that can be applied on top of any platform. Zero-trust security \cite{zerotrust} is a security model that assumes that any device or user is a potential threat, even if they are located within a secure network perimeter. Therefore, zero-trust security requires that every interaction between clients and servers is verified and authenticated before being processed, making it ideal for heterogeneous platform deployments. However, there must be a way to verify the identities of the clients to determine whether they should be granted access.

As previously stated, a colony is a collection of distributed executors. Executors part of the same colony trust each other and can thus submit function specifications and get process execution assignments. To implement such a scheme, the Colonies server must be able to verify the identity of the executors and then check their colony membership. This can be accomplished in various ways. One solution is to use public key encryption and assign each executor a pair of keys - a public key and a private key. The public key is openly available and uploaded to the Colonies server, whereas the private key is kept secret and is only used by the executor to sign API request messages. The Colony server can then verify the signature of incoming request messages and use the public keys to determine whether an executor is a member of a certain colony.

ECDSA (Elliptic Curve Digital Signature Algorithm) \cite{ecdsa} is a digital public key encryption signature algorithm. It is widely used in blockchains such as Bitcoin and Ethereum to verify transactions. One of the advantages of ECDSA is its ability to recover public keys from received messages and signatures without explicitly transmitting the public keys. This is particularly useful as the identity of an executor can be calculated simply as the SHA-3 hash of the recovered signature, saving space and eliminating the need for storing public keys. 

In ColonyOS, there exist three distinct roles, each with its own set of responsibilities and authority levels. The Colonies server owner holds the highest level of authority and has permission to add new colonies to the Colonies server. The colony owner is responsible for the management of a given colony and has permission to register or unregister executors within that colony. Executors and users, on the other hand, have the lowest level of authority and are only authorized to manage processes within their appointed colony. This functionality can be implemented by introducing a new table to the Colonies database, the colony table.

\begin{table}[h]
    \caption{Colony Id: \(4787a5071856a4acf702b2ffcea422e3237a679c681314113d86139461290cf4\)}
	\centering
	\begin{tabular}{ll}
		\toprule
		\cmidrule(r){1-2}
        Executor Id & Executor Name \\
		\midrule
        \(8a491bcac0be623a54411dd0934fdcdc9c844de5700527a7dbc9da08a6a8310d\) & $E_{1}$ \\
        \(6a65f40343999415f47e739f09b625ab437b069b5f591f9844c234533f505bee\) & $E_{2}$ \\
        \(41e4c1b10f92a53b7a5a86620ed366635901a1a96c48312a0ef566a13217fe03\) & $E_{3}$ \\
		\bottomrule
	\end{tabular}
	\label{coltable}
\end{table}

Table \ref{coltable} shows an example of a colony table consisting of three executors. In this case, to register a new executor, the SHA-3 hash of the recovered signature needs to match the identity \(4787a5071856a4acf702b2ffcea422e3237a679c681314113d86139461290cf4\) stored in the colony table. This operation can only be performed by the colony owner possessing the corresponding private key. Similarly, when an executor connects to the Colonies server to either submit a function specification or get a process assignment, it needs to sign the API message with its private key. The Colonies server will then recover the identity of the executor and look up the colony table to check if the executor is a rightful colony member. 

An advantage of using a stateless architecture for workflow management as proposed in the paper is the ability to monitor executor performance to detect any unusual or suspicious behaviors. Because the execution history is recorded in a database, it is possible to track the activity of individual executors and identify any unusual patterns. For example, if an executor suddenly starts to consume a lot of resources, it could indicate a performance issue or it could be a sign of malicious activity, such as a denial-of-service attack. Automatic monitoring can for example be implemented using anomaly detection algorithms and machine learning to grade the performance of each executor. Administrators can then be automatically notified if the grade of an executor exceeds a certain threshold, enabling them to take appropriate actions.

\section{Implementation}
ColonyOS is implemented in Golang and is publicly available on GitHub\footnote{https://github.com/colonyos/colonies} under the MIT license. It consists of a statically compiled binary that offers a CLI tool for deploying Colonies servers or managing the system. Furthermore, there are several software development kits (SDKs) available for various programming languages, including Golang\footnote{https://github.com/colonyos/colonies/tree/main/pkg/client}, Rust\footnote{https://github.com/colonyos/rust}, Julia\footnote{https://github.com/colonyos/Colonies.jl}, JavaScript\footnote{https://github.com/colonyos/colonies.js}, Python\footnote{https://github.com/colonyos/pycolonies}, and Haskell\footnote{https://github.com/colonyos/haskell}. As an introduction to ColonyOS, the following section gives a brief overview how to use the Python SDK.

\subsection{Python SDK}
A ColonyOS application typically consists of a set of executors written in different programming languages and deployed across multiple platforms. The executors are small microservices that interact with one another by submitting function specifications or workflows, which are then executed by other executors that are part of the same colony.

The first step in implementing an executor in Python is to create a private key and identity, followed by registering the executor with a colony. As mentioned before, executor can registration can only be carried out by the colony owner using the colony private key. Code in Listening \ref{code:regexecutor} illustrates how to register an executor, including registering functions a specific executor is capable of running. In the example, an executor of type \emph{helloworld\_executor} is registered to run a function named \emph{helloworld}.

\begin{lstlisting}[showstringspaces=false, frame=lines, numbers=left, numberstyle=\scriptsize, backgroundcolor=\color{background}, basicstyle=\small, label=code:regexecutor, language=Python, caption=Register an executor to a colony in Python.]
colonies = Colonies("localhost", 50080)

crypto = Crypto()
colonyid = "4787a5071856a4acf702b2ffcea422e3237a679c681314113d86139461290cf4"
colony_prvkey = "ba949fa134981372d6da62b6a56f336ab4d843b22c02a4257dcf7d0d73097514"
executor_prvkey = crypto.prvkey()
executorid = crypto.id(executor_prvkey)

executor = {
    "executorname": "helloworld_executor",
    "executorid": executorid,
    "colonyid": colonyid,
    "executortype": "helloworld _executor"
}

colonies.add_executor(executor, colony_prvkey)
colonies.approve_executor(executorid, colony_prvkey)

# register capability run the helloworld function
colonies.add_function(executorid, colonyid, "helloworld",  executor_prvkey)
\end{lstlisting}

After registration, the executor must connect to the Colonies server and request process assignments of the type \emph{helloworld\_executor}. Note that executor properties, such as the executor type, are not explicitly set in the \emph{assign} request, but rather derived implicitly by the Colonies server from the executor identity, which is derived from the signature generated by the executor's private key. This approach ensures that the executor can only be assigned processes that match its registered capabilities.

\begin{lstlisting}[showstringspaces=false, frame=lines, numbers=left, numberstyle=\scriptsize, backgroundcolor=\color{background}, basicstyle=\small, language=Python, label=code:assign, caption=Assigning and executing a process in Python.]
while True:
    process = colonies.assign(colonyid, 10, executor_prvkey)
    if process["spec"]["funcname"] == "helloworld":
       colonies.close(process["processid"], ["hello world"], executor_prvkey)
\end{lstlisting}

This \emph{assign} request is always initiated by the executors, allowing executors to be deployed behind firewalls. When receiving the request, the Colonies server hangs the request until a matching process is found or a timer expires. In the example in Listing \ref{code:assign}, the timer is set for 10 seconds, after which an exception is raised, and the executor must reconnect to the server.

When assigned to a process, the executor interprets the metadata stored in the process data structure, including the function name and arguments, and performs some kind of computation such as preprocessing data or training a neural network. Upon completion, the executor closes the process and sets the output. In the example, the output is set to the string \emph{helloworld}.

\begin{lstlisting}[showstringspaces=false, frame=lines, numbers=left, numberstyle=\scriptsize, backgroundcolor=\color{background}, basicstyle=\small, language=Python, label=code:submit, caption=Submitting a function specification.]
func_spec = create_func_spec(func="helloworld",
                             colonyid=colonyid,
                             executortype="helloworld_executor",
                             priority=0,
                             maxexectime=100,
                             maxretries=3)
process = colonies.submit(func_spec, executor_prvkey)
\end{lstlisting}

Listing \ref{code:submit} demonstrates how to submit a function specification using the Python SDK and call the \emph{helloworld} function. Note that the executor must complete the process within 100 seconds. The process is reassigned to another executor if the execution takes too long a time. The previous executor then receives an error when trying to send a \emph{close} request. As discussed in Section \ref{stateless_error_management}, establishing boundaries on computations is critical to achieving robustness and supporting infrastructural changes that may result from DevOps or IT operations.

\begin{figure}
     \centering
     \begin{subfigure}[b]{0.49\textwidth}
         \centering
         \includegraphics[scale=0.17]{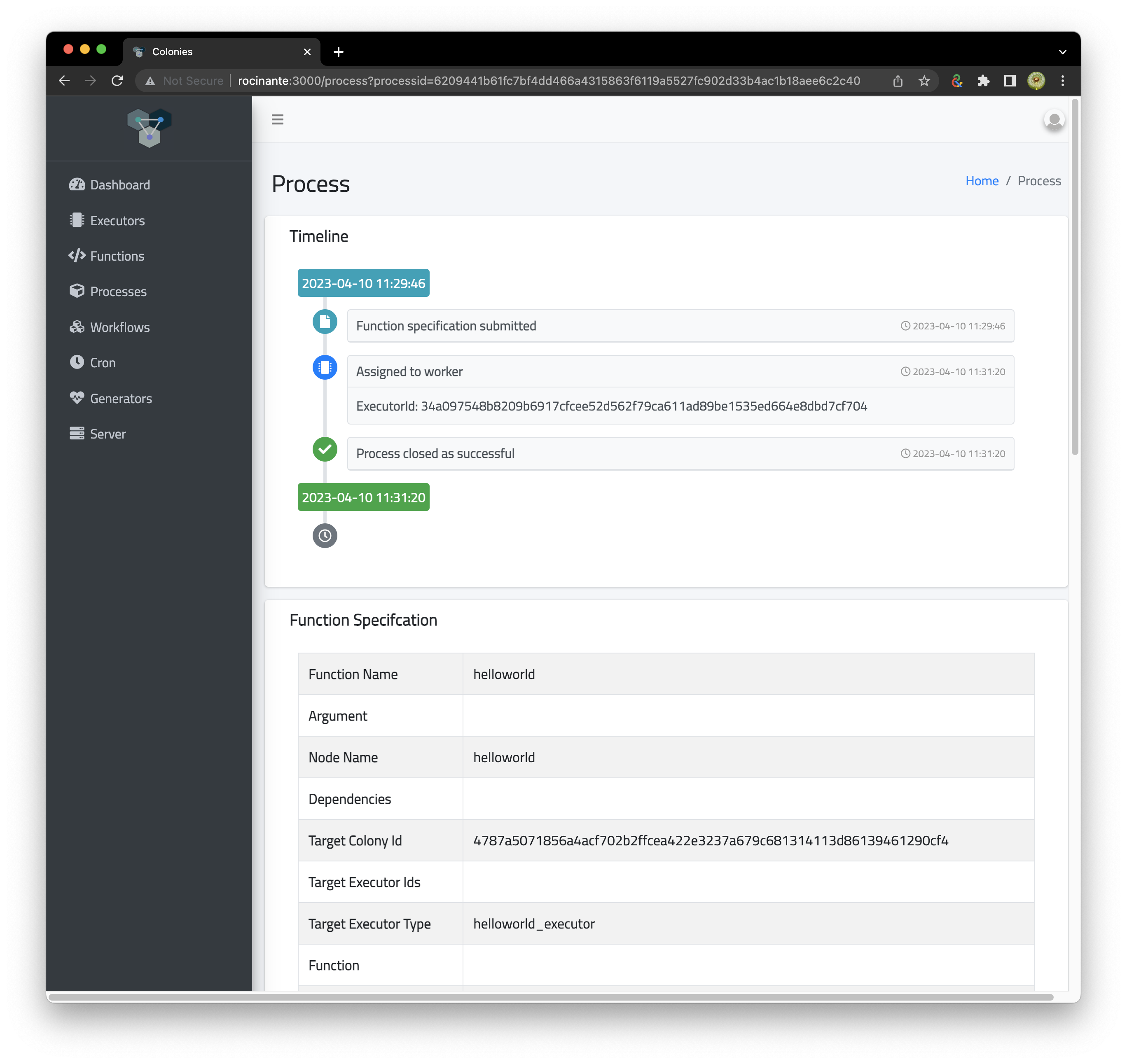}
         \caption{Screenshot of process information.}
     	 \label{fig:dashboard1}
     \end{subfigure}
     \hfill
     \begin{subfigure}[b]{0.49\textwidth}
         \centering
         \includegraphics[scale=0.17]{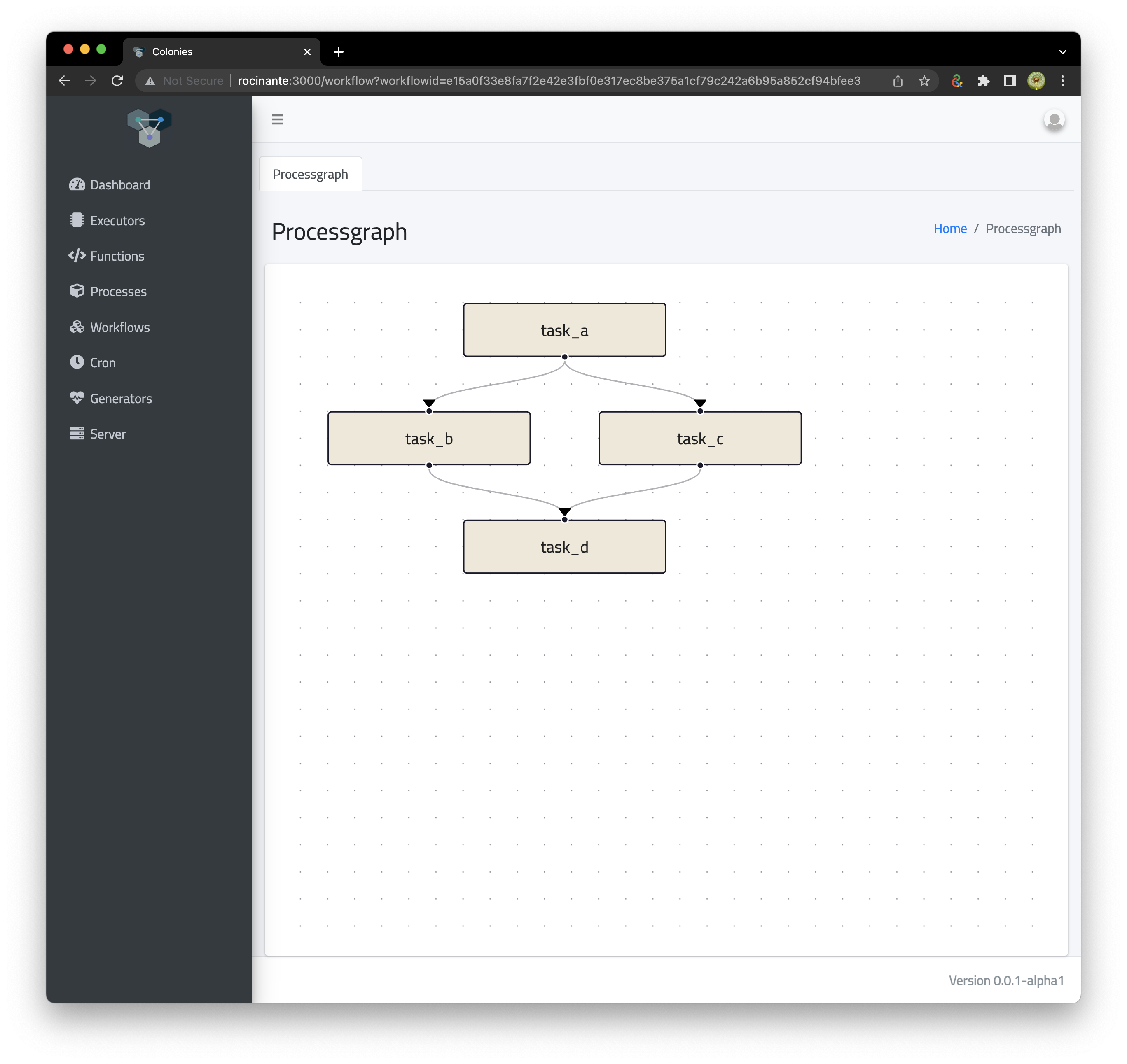}
      	 \caption{Screenshot of a workflow.}
     	 \label{fig:dashboard2}
     \end{subfigure}
     \caption{Screenshot of the Colonies Dashboard tool.}
\end{figure}

% \begin{figure}[h]
% 	\centering
%     \includegraphics[scale=0.23]{dashboard1.png}
% 	\caption{Process information. The timeline makes it possible to follow the execution of process in realtime.}
% 	\label{fig:dashboard1}
% \end{figure}

Figure \ref{fig:dashboard1} shows a screenshot of the ColonyOS dashboard tool, which enables users to monitor the status of processes and executors. The timeline depicted in the figure allows for real-time monitoring of process execution, similar to tracking a physical package sent through a postal service. As previously discussed, it's also possible to subscribe to a process and receive information about its progress, which can be used as another point of integration. For example, sending a notification on Slack if a process fails.

\subsection{Workflows}
A workflow is just a set of processes with dependencies, which is represented as a DAG internally in Colonies. Workflows can either be submitted using SDKs or the Colonies CLI, which supports submission of JSON files. Listening \ref{code:workflow_spec} shows an example of a workflow consisting of four processes, each requiring different types of executors. The management of workflows is entirely handled by Colonies, and each executor executes only processes that are part of the graph for which it is responsible for. This approach significantly reduces the complexity of designing, implementing and operating large-scale AI applications and data pipelines. 

\begin{lstlisting}[basicstyle=\small, label=code:workflow_spec, language=json, basicstyle=\small, caption=Example of a workflow expressed in JSON.]
[ { "nodename": "task_a",
    "funcname": "echo",
    "conditions": { "executortype": "executor_type1", "dependencies": [] } },
  { "nodename": "task_b",
    "funcname": "echo",
    "conditions": { "executortype": "executor_type2", "dependencies": ["task_a"] } },
  { "nodename": "task_c",
    "funcname": "echo",
    "conditions": { "executortype": "executor_type3", "dependencies": ["task_a"] } },
  { "nodename": "task_d",
    "funcname": "echo",
    "conditions": { "executortype": "executor_type4", "dependencies": ["task_b", "task_c"] }
  } ]
\end{lstlisting}

% \begin{figure}[h]
% 	\centering
%     \includegraphics[scale=0.23]{dashboard2.png}
% 	\caption{Screenshot of a workflow in the Colonies dashboard.}
% 	\label{fig:dashboard2}
% \end{figure}

In addition to monitoring processes and executors, the ColonyOS dashboard can be used to visualize workflows. Figure \ref{fig:dashboard2} shows a DAG created when submitting the workflow code shown in Listing \ref{code:workflow_spec}. This functionality makes it possible to follow workflow execution progress.

\section{Discussion}
This paper has presented an open-source meta-operating system called ColonyOS designed to streamline execution of computational workloads across platforms. Instead of directly connecting different systems together, ColonyOS enables special kinds of microservices so-called executors to publish instructions which are subsequently assigned to other executors. By chaining instructions together, it becomes possible to execute workloads that can operate seamlessly across different platforms. 

ColonyOS builds on the principle: \emph{security first}. Public key encryption enables administrators and users to maintain strict control over a colony, even when executors are dispersed throughout the Internet. Each colony is governed by a private key that serves as a central authority, analogous to \emph{one ring to rule them all}, making it possible to control a colony as if it was a single machine. The zero-trust principle, \emph{trust, but verify} also enables development of immune system-like functionality where malfunctional executors can be excluded and replaced with fresh new ones, ensuring that the overall performance and reliability of the system is not compromised. 

A significant portion of the paper has focused on describing how to handle resilience to allow software components to fail without compromising the entire system. A critical design decision that makes ColonyOS robust and scalable is its stateless architecture. ColonyOS also makes sure that workflow execution runs to completion, even if some executors fail. This makes it possible to implement robust workflows, but also efficiently handle IT operations such as Continuous Integration/Continuous Delivery (CI/CD). Both security and resilience are crucial factors to consider when developing an AI workflow management system that is capable of functioning across platforms and handling constantly changing infrastructures. 

It has been mentioned in the paper several times that ColonyOS is designed to simplify integration with other systems. Microservices is a design principle used by ColonyOS to simplify integration. As executors are designed to be loosely-coupled microservices, the entire system becomes easier to develop, test, and maintain. For example, different teams can take on responsibility to develop a certain executor, which can then be used as a part of complex workflows running across platforms. By designing each executor as a standalone microservice, it also becomes easy to scale the system by simply deploying more executors. This also opens up for an intriguing possibility to delegate colony management responsibility to a software agent that can automatically provision and deploy executors on demand.

While microservices is a powerful design pattern, ColonyOS goes beyond the conventional microservice model and offers a distributed microservice architecture allowing microservices to reside anywhere on the Internet. By acting as an intermediary layer, ColonyOS provides a unified interface to facilitate interaction between these microservices. This makes it possible to deploy executors behind firewalls, and even run on an executor in a web browser.

With ColonyOS, it becomes possible to develop hyper-distributed applications that can run AI workloads across multiple platforms. For example, it becomes possible to train a ML model on an HPC system, then use CFS to synchronize the trained model to a cloud environment, seamlessly updating a production pipeline. Additionally, ColonyOS makes it possible to scale on-premises clusters with cloud resources, allowing for computational scaling on demand. This capability also makes it possible to optimize the use of cloud resources, for example minimizing cloud resources to lower costs or energy utilization. 

In summary, ColonyOS aims to provide continuous access to computing resources, enabling seamless execution and data flow across various platforms. This paper has proposed a meta-operating system to navigate the complexities of a computing continuum composed of heterogeneous platforms, positioning it as an ideal solution for managing next-generation AI services. Consequently, ColonyOS represents a significant step toward the realization of a fully integrated computing ecosystem.

\bibliographystyle{IEEEtran.bst}
\bibliography{references} 

\end{document}